\documentclass[11pt,fleqn]{article}
\usepackage[utf8]{inputenc}
\usepackage{graphicx}
\usepackage{float}
\usepackage{amssymb,amsmath,amsfonts, mathtools}
\usepackage{enumerate}
\usepackage{xcolor}

\definecolor{hyperref}{RGB}{026,028,185}
\usepackage[bookmarks=true,colorlinks=true,linkcolor=hyperref,citecolor=hyperref,urlcolor=hyperref,bookmarksnumbered]{hyperref}
\usepackage{footmisc}

\def\clock{{\count0=\time
		\divide\count0 60
		\ifnum\count0<10 0\fi\the\count0
		\multiply\count0 -60 \advance\count0 \time
		:\ifnum\count0<10 0\fi \the\count0
}}
\newcommand{\timestamp}{{\small\vbox{\hbox{\tt\jobname.tex}
			\hbox{\the\day/\the\month/\the\year, \clock}}}}

\newcommand{\ba}{\begin{eqnarray}}
\newcommand{\ea}{\end{eqnarray}}

\renewcommand{\deg}{\text{deg}\,}
\renewcommand{\dim}{\text{dim}\,}

\newcommand{\be}{\begin{equation}}
\newcommand{\ee}{\end{equation}}

\makeatletter
\let\old@startsection=\@startsection
\let\oldl@section=\l@section
\renewcommand{\@startsection}[6]{\old@startsection{#1}{#2}{#3}{#4}{#5}{#6\mathversion{bold}}}
\renewcommand{\l@section}[2]{\oldl@section{\mathversion{bold}#1}{#2}}
\makeatother

\numberwithin{equation}{section}

\usepackage{color}

\def \adss {$AdS_5 \times S^5$\ }

\newcommand{\beq}{\begin{equation}}
\newcommand{\eeq}{\end{equation}}

\begin{document}
	\renewcommand{\thefootnote}{\arabic{footnote}}
	
	\overfullrule=0pt
	\parskip=2pt
	\parindent=12pt
	\headheight=0in \headsep=0in \topmargin=0in \oddsidemargin=0in
	
	\vspace{ -3cm} \thispagestyle{empty} \vspace{-1cm}
	\begin{flushright} 
		\footnotesize
		HU-EP-22/01-RTG
	\end{flushright}%

	\begin{center}
		\vspace{1.2cm}
		{\Large\bf \mathversion{bold}
			 Lattice perturbation theory for the null cusp string 
		}
		
		\author{Gabriel~Bliard\thanks{XYZ} \and DEF\thanks{UVW} \and GHI\thanks{XYZ}}
		\vspace{0.8cm} {
			Gabriel~Bliard$^{~a,b,}$\footnote{{\tt $\{$gabriel.bliard,ilaria.costa,patella$\}$@\,physik.hu-berlin.de}}, Ilaria~Costa$^{a,1}$, 
			Valentina~Forini$^{c,a}$\footnote{\tt valentina.forini@city.ac.uk}, Agostino~Patella$^{a,1}$}
		\vskip  0.5cm
		
		\small
		{\em
			$^{a}$Institut f\"ur Physik, Humboldt-Universit\"at zu Berlin, IRIS Adlershof, \\Zum Gro\ss en Windkanal 2, 12489 Berlin, Germany  
			\vskip 0.05cm
			$^{b}$      Dipartimento di Fisica, Universit\`a di Parma,\\ Viale G.P. Usberti 7/A, 43100 Parma, Italy, 
			\vskip 0.05cm
			$^{c}$      Department of Mathematics, City, University of London\\
			Northampton Square, EC1V 0HB London, United Kingdom }
		\normalsize
		
	\end{center}

	\vspace{0.3cm}
	\begin{abstract} 
		We reconsider the problem of discretising the worldsheet for the gauge-fixed Green-Schwarz superstring on a null cusp background, and present a setup which fully preserves its global $U(1)\times SU(4)$ symmetry.   We discuss divergences by power counting on the lattice, and study renormalizability at one loop with the example of one-point functions and one bosonic correlator of the worldsheet excitations.  In order to remove  UV divergences at one loop, it is necessary to introduce two extra parameters in the action, which need to be either fine-tuned at tree level or renormalized at one-loop.
	\end{abstract}
	
	\newpage
	
   \setcounter{footnote}{0}
   
	\tableofcontents

	\section{Introduction and discussion}

In the framework of the AdS/CFT~\cite{Maldacena:1998im,Rey:1998ik}
correspondence, the expectation value of a light-like cusped Wilson loop in
$\mathcal{N}=4$ super Yang-Mills is equal to the partition function of an open
string propagating in AdS$_5\times S^5$ space and ending on the loop at the AdS
boundary. In practice one writes
\begin{equation}
\label{cusp_anom}
\left\langle\mathcal{W}_{\text{cusp }}\right\rangle
=
\int \mathcal{D} Y \mathcal{D} \Psi \, e^{-S_{\text{cusp}}(X_\text{cl} + Y,\Psi)} \equiv e^{-\frac{f(g)}{8} V_{2}}
\, ,
\end{equation}
where $S_{\text{cusp}}$ is obtained from the Green-Schwarz AdS$_5\times S^5$
superstring action, by parametrizing the fluctuations of the bosonic degrees of
freedom $X = X_\text{cl} + Y$ around the classical null-cusp solution
$X_{\mathrm{cl}}$~\cite{Gubser:2002tv,Kruczenski:2002fb}, and by fixing the
local bosonic (diffeo) and fermionic (kappa) symmetries e.g. to light-cone
gauge~\cite{MT2000}. The free energy of the open string is proportional to the
worldsheet volume $V_2$ and we refer to the prefactor $f(g)$  as the
\textit{cusp anomaly}~\footnote{
In some literature, $f(g)$ is called ``scaling function''. From the gauge theory
point of view, it governs the logarithmic behavior in the large spin anomalous
dimensions of twist-two operators, and equals twice the cusp anomalous dimension
of light-like Wilson loops~\cite{Belitsky:2006en}. The same can been
seen~\cite{Kruczenski:2007cy} at the level of the dual classical string solutions,
respectively~\cite{Gubser:2002tv} and~\cite{Kruczenski:2002fb}. The
normalization factor $1/8$ in~\eqref{cusp_anom} also takes into account the
conventions of~\cite{Giombi}. \label{fn1}%
}~\cite{Korchemsky:1987wg,Korchemsky:1992xv,Belitsky:2006en}. The cusp anomaly
is a function of the coupling constant
$g=\frac{R^2}{4\pi\alpha'}=\frac{\sqrt{\lambda}}{4\pi}$, where $R$ is the common
radius of AdS$_5$ and $S^5$, $\alpha'$ is the squared string scale, while
$\lambda$ is the 't Hooft coupling on the gauge side of the AdS/CFT
correspondence. The cusp anomaly has been calculated to next-to-next-to-leading
order in a perturbative expansion in $g^{-1}$~\cite{Giombi} and in dimensional
regularization. Assuming integrability~\cite{BES,Beisert:2010jr} and using the
corresponding
technology~\cite{BES,Basso:2013vsa,Basso:2013aha,Fioravanti:2015dma}, the cusp
anomaly can be evaluated also at finite coupling.

The Green-Schwarz AdS$_5\times S^5$ string is expected to be defined also at the
non-perturbative level. A valid question is whether the non-perturbative regime
of the $\sigma$-model, which describes the AdS$_5\times S^5$ string at tree-level
in string perturbation theory, is accessible through a lattice discretization of
the worldsheet (while target space remains continuous). This question is
motivated by the success of the lattice as a UV non-perturbative regulator of
Quantum Chromodynamics. This approach has been pioneered
in~\cite{Roiban,Forini:2016gie,Bianchi:2019ygz, Forini:2021keh}, where a
lattice-discretized version of $S_{\text{cusp}}$ has been introduced and also
used to perform of Monte Carlo simulations~\footnote{ Other lattice approaches
to AdS/CFT include~\cite{Catterall_physrept,
Catterall:2014vka,Schaich:2014pda,Catterall:2014vga,Catterall:2015ira,
Schaich:2015ppr,Schaich:2015daa,Giedt:2016yfw,Catterall:2017lub,Jha:2017zad,Rinaldi:2017mjl,Hanada:2018qpf,
Catterall:2020nmn,Watanabe:2020ufk,Gharibyan:2020bab,Bergner:2021ffz,
Dhindsa:2021irw,Bergner:2021goh}, see also~\cite{Schaich:2018mmv} and references
therein.}.

Once a lattice discretization of $S_{\text{cusp}}$ and of the path integral is
proposed, one still needs to understand whether the continuum limit (i.e. the
limit in which the lattice spacing $a$ vanishes) exists for physical
observables, and whether the obtained continuum theory has the desired defining
properties. Notice that the inverse lattice spacing $a^{-1}$ is nothing but a UV
cutoff, and the question of the existence of the continuum limit is logically
equivalent to the question of cancellation of UV divergences after
renormalization: \textit{once a discretized action is defined as a function of
a finite number of bare parameters, is it possible to cancel all UV divergences
in on-shell observables with a redefinition of the bare parameters?}
The existence of the continuum limit at the non-perturbative level
is a very complicated issue, both theoretically and numerically. However, if the
lattice regularization makes sense at all, then one should recover the correct
continuum theory also order-by-order in the perturbative expansion, i.e. in
powers of $g^{-1}$. The goal of this paper is precisely to set the stage for
such a perturbative expansion, and to discuss some peculiarities of the lattice
regulator.

In Section~\ref{sec:discretization}, we present a new discretization for
$S_{\text{cusp}}$. Contrarily to the actions proposed and used
in~\cite{Bianchi:2016cyv,Bianchi:2019ygz,Forini:2017mpu}, the new action is
invariant under the full $U(1)\times SU(4)$ group of internal symmetries. As
usual in QFT, more symmetries mean less UV divergences. In
Section~\ref{sec:expansion}, we parametrize the fluctuations around the classical
solution in analogy to what is usually done in the continuum~\cite{Giombi} and we
calculate the propagators for the lattice discretized theory.

In Section~\ref{sec:powercounting} we calculate the superficial degree of
divergence of the generic Feynman diagram and we show that power counting
suggest that infinitely many counterterms are needed at every order in the
perturbative expansion to cancel all UV divergences. 

This result is not so surprising, as the Green-Schwarz action expanded around a
classical background is known to be formally power-counting
non-renormalizable~\cite{Polyakov:2004br,Roiban:2007jf,Roiban:2007dq}. However
in the continuum, when using the regularization introduced
in~\cite{Roiban:2007jf,Roiban:2007dq} to which we refer as ``dimensional
regularization'' in what follows, the cusp anomaly turns out to be finite
without any counterterm, at least up to two loops~\cite{Roiban:2007dq,Giombi}.
The cancellation of divergences has been verified similarly for the two-point
functions and the dispersion relation of excitations near a long spinning string
in AdS$_5$ at one loop~\cite{Giombi:2010bj}, and for a  ``generalized scaling
function'' governing the energy of a string spinning both in AdS$_5$ and in
$S^5$ at two loops~\cite{Giombi:2010fa,Giombi:2010zi}~\footnote{The classical
worldsheet theory of the long spinning string in AdS$_5$ is equivalent, via an
analytic continuation and a global conformal transformation, to that of the
light-like cusp solution which is of interest here, see
footnote~\footref{fn1}.}.

In order to understand whether similar cancellations of UV divergences happen
also in the lattice discretized theory, we calculate the cusp anomaly, the
one-point function of the field $\phi$ (which parametrizes the radial direction
of AdS$_5$), and the two-point function of $x$, which parametrizes the
fluctuations of the string at the AdS$_5$ boundary. These calculations are
presented in Section~\ref{Sec: calculations}. We will see explicitly that, in
the considered lattice discretization, the situation is quite more complicated
than in dimensional regularization, and it is related to the presence of power
divergences. We observe the following interesting facts:
\begin{enumerate}
   \item The quadratic divergences cancel at one loop in the one-point function
   of $\phi$ and in the two-point function of $x$ (while they are subtracted by
   hand in the cusp anomaly). At one loop, these cancellations seems quite
   robust in the sense that they will always happen in any reasonable
   discretization of the action.
   \item Linear divergences arise as well, and they generally do not cancel in
   all considered observables. These divergences are very specific of the
   lattice discretization, and arise from the particular choice of forward and
   backward discrete derivatives. In order to cure this problem we have
   introduced two extra parameters $b_\pm$ in the action that would be naturally
   set to $1$ at the classical level. In order to remove the linear divergences
   at one loop, these parameters need to be either fine-tuned at tree level or
   renormalized at one-loop.
   \item Once the linear divergences are removed by tuning or renormalization,
   the logarithmic divergences cancel in the cusp anomaly and in the two-point
   function of $x$ (while they survive in the one-point function of $\phi$ in
   analogy to the continuum). Moreover the continuum limit of the cusp anomaly
   and of the dispersion relation of the worldsheet excitation with the quantum
   numbers of the field $x$ are the same as the ones obtained in dimensional
   regularization.
\end{enumerate}

The extra parameters $b_\pm$ do not seem to have any deep meaning besides the
fact that they make the bare propagators particularly simple. Moreover we do not
claim that the introduction and fine-tuning of these two extra parameters is
enough to make all physical observables finite at all orders in perturbation
theory, and this is in fact highly unlikely. Still, one would like to understand
whether the number of parameters needed to achieve finiteness of physical
observables via fine-tuning or renormalization is finite or not. If infinitely
many parameters are necessary, then the discretized model has no predictivity,
and it cannot be used as a viable non-perturbative definition of the
AdS$_5\times S^5$ string in null-cusp background. A complete one-loop analysis
of the divergences of $n$-point functions may help shed light on this issue, and
we plan to carry it on in the future, with the technology developed in this
paper.

One may also try to find a general mechanism that prevents linear divergences in
the first place. Building on the idea that odd powers of $a$ must be accompanied
by odd powers of $m$, one may try to exploit a spurionic symmetry that involves
the replacement $m \to -m$, the reflection of both worldsheet coordinates and an
$SO(5)$ rotation, which is enjoyed by the continuous action. Such spurionic
symmetry is broken by our lattice discretization. Some preliminary explorations
that we do not report here indicate that it is not completely trivial to
preserve this symmetry on the lattice while avoiding the doubling problem.
Different options in this direction will be explored in the future.


%

%

	\section{$U(1)\times SU(4)$ invariant discretization}
\label{sec:discretization}

In the continuum, the \adss superstring action  in a AdS-lightcone
gauge-fixing describing quantum  fluctuations around  the null-cusp background
reads~\cite{Giombi} 
\begin{eqnarray}
\nonumber
S^\text{cont}_{\rm cusp}=g \int dt ds&& \Bigg\{ 
\left| \partial_t x + \tfrac{m}{2} x \right|^2
+ \tfrac{1}{z^4} \left| \partial_s x - \tfrac{m}{2} x \right|^2
+ \left( \partial_t z^M + \tfrac{m}{2} z^M + \tfrac{i}{z^2} z_N \eta_i \left(\rho^{MN}\right)_{\phantom{i}j}^{i} \eta^j \right)^2
\\ \nonumber &&
+ \tfrac{1}{z^4} \left( \partial_s z^M - \tfrac{m}{2} z^M \right)^2  
+i\, \left( \theta^i \partial_t \theta_i + \eta^i \partial_t \eta_i + \theta_i \partial_t \theta^i + \eta_i \partial_t \eta^i \right) - \tfrac{1}{z^2} \left( \eta^i \eta_i \right)^{2}  \\ \nonumber &&  
+2i \, \Big[
\tfrac{1}{z^3} z^M \eta^i \left( \rho^M \right)_{ij}
\left( \partial_s \theta^j - \tfrac{m}{2} \theta^j - \tfrac{i}{z} \eta^j \left( \partial_s x -\tfrac{m}{2} x \right) \right)
\\ && \qquad
+ \tfrac{1}{z^3} z^M \eta_i \big( {\rho^M}^\dagger \big)^{ij}
\left( \partial_s \theta_j - \tfrac{m}{2} \theta_j + \tfrac{i}{z} \eta_j \left( \partial_s x -\tfrac{m}{2} x \right)^* \right)
\Big]
\,\Bigg\}
\, ,
\label{S_cusp_cont}
\end{eqnarray}
where
\begin{itemize}
	\item  $x$ is a complex bosonic field whose real and imaginary part
	parametrize the fluctuations of the string (in light-cone gauge) at the
	boundary of AdS$_5$. 
	\item $z^M$ are six real bosonic field, i.e. $M=1,\cdots, 6$; $z=\sqrt{z^M
	z^M}$ is the radial coordinate of the AdS$_5$ space, while $u^M = z^M/z$
	identifies points on $S_5$.
	\item the Gra\ss mann-odd fields $\theta^i = (\theta_i)^\dagger,~\eta^i =
	(\eta_i)^\dagger, \, i=1,2,3,4$ are complex anticommuting variables (no
	Lorentz spinor indices appear);
	\item the matrices $(\rho^{MN})_i^{\hphantom{i} j} = (\rho^{[M} \rho^{\dagger
	N]})_i^{\hphantom{i} j}$ are the $SO(6)$ generators.  $\rho^{M}_{ij}
	$\footnote{By convention, we will write the indices of $\rho$ as down and
	those of $\rho^\dag$ as up.}  are the (traceless) off-diagonal	blocks of
	$SO(6)$ Dirac matrices $\gamma^M$ in  chiral representation, see
	Appendix~\ref{app:discretization}. 
\end{itemize}

The massive parameter $m$ keeps track of the (dimensionful) light-cone momentum $P_+$, set to one in~\cite{Giombi}. The action~\eqref{S_cusp_cont} is invariant under a $U(1)\times SU(4)$ global symmetry defined by
\begin{gather}
   z^M \to \text{Ad}(U)^{MN} 
   z^N \ , \quad \theta^i \to U^i_{\phantom{i}j} \theta^j \ , \quad \eta^i \to U^i_{\phantom{i}j} \eta^j \ , \\
   x \to e^{i \alpha} x \ , \quad \theta^i \to e^{i \alpha/2} \theta^i \ , \quad \eta^i \to e^{-i \alpha/2} \eta^j \ ,
\end{gather}
%
where $U$ is an element of $SU(4)$ and its representative in the adjoint, $\text{Ad}(U)$, is an element of $SO(6)$. While the original Green-Schwarz AdS$_5\times S^5$ string action is invariant under diffeomorphisms and $\kappa$-symmetry, these local symmetries have be fixed by the choice of light-cone gauge in eq.~\eqref{S_cusp_cont}. Notice that the action is not invariant under worldsheet rotations, parity ($s \rightarrow -s$), or time reversal ($t\rightarrow -t$).

\bigskip

In order to define the lattice-discretized theory we need to provide a
discretized action, but also an explicit expression for the measure. We choose
to use a flat measure for the fields, but we keep in mind that this choice is
quite arbitrary as it is not invariant under reparametrization of the target
AdS$_5 \times S_5$ target space. Given a generic observable $A$, expectation values in the lattice discretized theory are defined by
\begin{equation}
\label{path_int_latt}
   \langle A \rangle = \frac{1}{Z_{\text{cusp}}} \int dx dx^* d^6z d^4\theta d^4\theta^\dag d^4\eta d^4\eta^\dag
   \, e^{-S_{\text{cusp}}} A
   \, ,
\end{equation}
where $d f \equiv \prod_{s,t} d f(s,t)$, as usual the partition function $Z_{\text{cusp}}$ is fixed by the requirement $\langle 1 \rangle = 1$,  and $S_{\text{cusp}}$ refers now to the discretised action, that we choose to be 
\begin{eqnarray}\nonumber
\!\!\!\!\!\!\!\!\!	
S_{\rm cusp}=g
	\sum_{s, t} a^2&&\!\!\! \!\!\Bigg\{ \!\!
	\left| b_+ \hat\partial_t x + \tfrac{m}{2} x \right|^2
  \!\! + \tfrac{1}{z^4}  \left| b_- \hat\partial_s x - \tfrac{m}{2} x \right|^2
	\!\!+ \big( b_+ \hat\partial_t z^M + \tfrac{m}{2} z^M + \tfrac{i}{z^2} z^N \eta_i (\rho^{MN})_{\phantom{i}j}^i \eta^j \big)^2
   \\ \nonumber && 
	+ \tfrac{1}{z^4} \big( \hat\partial_s z^M \hat\partial_s z^M + \tfrac{m^2}{4} z^2 \big)
	+ 2 i\, \big( \theta^i \hat\partial_t \theta_i + \eta^i \hat\partial_t \eta_i \big)
   - \tfrac{1}{z^2} \left( \eta^i \eta_i \right)^2  \\\nonumber
	&& 
	+ 2i\, \Big[ \tfrac{1}{z^3} z^M \eta^i \left(\rho^M\right)_{ij}
	\big( b_+ \bar\partial_s \theta^j - \tfrac{m}{2} \theta^j - \tfrac{i}{z} \eta^j \big( b_- \hat\partial_s x - \tfrac{m}{2} x \big) \big)
	\\\label{S_cusp_latt} && \qquad
	+ \tfrac{1}{z^3} z^M \eta_i \big({\rho^M}^\dagger\big)^{ij} \big( b_+ \bar\partial_s \theta_j - \tfrac{m}{2} \theta_j + \tfrac{i}{z} \eta_j \big( b_- \hat\partial_s x^* -\tfrac{m}{2} x^* \big) \!\big)\!
   \Big]\Bigg\}
   \, .
\end{eqnarray}
The action is written in terms of the forward and backward discrete derivatives
\begin{equation}
\hat{\partial}_\mu f(\sigma)\equiv\frac{f\left(\sigma +a e_\mu\right)-f\left(\sigma\right)}{a}
\, , \qquad 
\bar{\partial}_\mu f(\sigma)\equiv\frac{f\left(\sigma\right)-f\left(\sigma -a e_\mu\right)}{a}
\,
\end{equation}
where $e_\mu$ is the unit vector in the direction $\mu=0,1$, and  $\sigma$ is a shorthand notation for $(s,t)$.

Notice that the proposed discretized action~\eqref{S_cusp_latt} depend on four
parameters: $g$, $m$, and the auxiliary parameters $b_\pm$. It is
straightforward to see that the discretized action $S_{\rm cusp}$ reduces to the
desired continuum action $S^\text{cont}_{\rm cusp}$ in the naive $a \to 0$ limit,
if $b_\pm \to 1$. However, as we will discuss in detail, the naive choice $b_\pm
= 1$ produces undesired UV divergences at one loop. The values of $b_\pm$ need
to be tuned in such a way that these UV divergences cancel. This is a sign of
the fact that the lattice regulator does not manage to reproduce the
cancellation of UV divergences that occurs in dimensional regularization.

An important feature of the proposed discretized action and measure it that they
are invariant under the full $U(1)\times SU(4)$ internal symmetry group. This is
in contrast to the discretization previously presented
in~\cite{Bianchi:2019ygz}. The key ingredient is the use of forward and backward
discrete derivatives for both the bosonic and the fermionic part of the action.
This is normally avoided for fields that satisfy first-order equations of motion
(usually fermions), since it breaks parity and time-reversal.  In our case, this
is not an issue because these symmetries are already broken in the continuum
action. In~\cite{Bianchi:2019ygz}, instead, the symmetric derivative was used
and, as in lattice QCD, a Wilson-like term was included to cure the resulting
doubling problem, while breaking either the $U(1)$ or the $SU(4)$ symmetry.

\section{Perturbative expansion}
\label{sec:expansion}

On the lattice as in the continuum, the perturbative series is obtained by expanding the action around of its minima. The $SU(4)$ symmetric point (all fields vanish in this point) is a singularity for the action because of the terms proportional to inverse powers of the radial coordinate $z$. As a consequence the minimum of the action must spontaneously break the internal symmetry. In the continuum an absolute minimum of the action is given by $x=x^*=0$ and $z^M = \delta^{M6}$, and any other absolute minimum is obtained by acting with the $SU(4)$ symmetry. One can easily check that these minima are also relative minima for the discretized action. We parametrize the fluctuations around the chosen minimum is the same way as it is done in the continuum~\cite{Giombi}
\begin{equation}\!\!\!\!\!\!\!\!\!\!\!\!\!
z=e^\phi\,,\qquad
   z^{a} = e^\phi \frac{y^a}{1+\frac{1}{4}y^2}
   \, , \qquad
   z^6 = e^\phi \frac{1-\frac{1}{4}y^2}{1+\frac{1}{4}y^2}
   \, , \qquad
   y^2 = \sum_{a=1}^5 (y^a)^2\,,\quad a=1,\dots,5
   \, .
\end{equation}
In terms of the new variables $\phi$ and $y^a$, the path-integral measure over
the $z^M$ fields reads
\begin{gather} 
  \prod_{M=1}^6 dz^M = 
   e^{\sum_{s,t} \left\{ 6\phi+ 5\log\left( 1+\frac{y^2}{4} \right) \right\}}
   \,
   d\phi \prod_{a=1}^5 d y^a
   \, .
\end{gather}
The contribution of the Jacobian determinant above can be conveniently included
in the effective action
\begin{gather}
   S_{\text{eff}} = S_{\text{cusp}} - \sum_{s,t} \left\{ 6\phi+ 5\log\left( 1+\frac{y^2}{4} \right) \right\}
   \, ,
   \label{eq:Seff}
\end{gather}
in terms of which expectation values of observables read
\begin{equation}
   \langle A \rangle = \frac{1}{Z_{\text{eff}}} \int dx dx^* d\phi d^5y d^4\theta d^4\theta^\dag d^4\eta d^4\eta^\dag
   \, e^{-S_{\text{eff}}} A
   \, .
\end{equation}
Notice that the sum in the contribution to the effective action of the Jacobian
determinant does not come with the corresponding $a^2$ factor, which means that
in the naive continuum limit it diverges like $a^{-2}$. This should be not
surprising: in the continuum this term would be proportional to $\delta^2(0)$
which yields a quadratic divergence in a hard-cutoff regularization (but it is
set to zero in dimensional regularization).

The perturbative expansion, i.e. the expansion in powers of $g^{-1}$, is
obtained by splitting the action $S_{\text{eff}} = S_0 + S_{\text{int}}$, where
$S_0$ contains all quadratic terms in the fields with a coefficient proportional
to $g^{-1}$, and $S_{\text{int}}$ contains all other terms. Notice that
$S_{\text{int}}$ also contains $g$-independent quadratic terms which comes from
the expansion of the Jacobian determinant. We focus here on the leading-order
quadratic action
\begin{eqnarray}
\nonumber
S_0 =
g \,a^2
\sum_{s, t}
\Bigg\{ &&
\left| b_+ \hat\partial_t x + \tfrac{m}{2} x\right|^2
+ \left| b_- \hat\partial_s x - \tfrac{m}{2} x \right|^2
\\ \nonumber &&
+ b_+^2 (\hat\partial_t y^a)^2 + m b_+ y^a \hat\partial_t y^a
+ (\hat\partial_s y^a)^2
\\ \nonumber &&
+ b_+^2 (\hat\partial_t \phi)^2 + m b_+ \phi \hat \partial_t \phi
+ (\hat\partial_s \phi)^2 + m^2 \phi^2
+2 i\left( \theta^i \hat{\partial}_t \theta_i +\eta^i \hat{\partial}_t \eta_i \right)
\\ &&
+2 i \eta^i (\rho^6)_{ij} \left( b_+ \bar{\partial}_s \theta^j - \tfrac{m}{2} \theta^j
\right)
+2 i \eta_i ({\rho^6}^\dagger)^{ij} \left( b_+ \bar{\partial}_s \theta_j - \tfrac{m}{2} \theta_j \right)
\Bigg\}
\, .
\label{S0}
\end{eqnarray}

The propagators are conveniently constructed by going in momentum space. Given a
function $f(s,t)$ in coordinate space, we denote by $\tilde{f}(p_0,p_1)$ the
corresponding function in momentum space. On the lattice, the two are related by
\begin{gather}\label{Eq:fourier}
	f(s,t) =
   \int_{-\pi/a}^{\pi/a} \frac{d^2p}{(2\pi)^2}e^{ip_0 t+ip_1s}\tilde{f}(p_0,p_1)
   \, , \qquad
   \tilde{f}(p_0,p_1)
   =
   \sum_{s,t} a^2 \, e^{-ip_0 t-ip_1s} f(s,t)
   \, .
\end{gather}
The function $\tilde{f}(p_0,p_1)$ is periodic in both components with period
$2\pi/a$, and momentum integrals are always restricted to $-\pi/a < p_k < \pi/a$
which shows explicitly that the lattice effectively enforces a hard cutoff in
momentum space. As in the continuum,  discrete derivatives  are diagonalized in
Fourier space,  and read
\begin{gather}
\widetilde{ \hat{\partial}_\mu f}(p_0,p_1) = i \hat{p}_\mu  \tilde{f}(p_0,p_1) 
\, , \qquad
\widetilde{ \bar{\partial}_\mu f}(p_0,p_1) = i \hat{p}_\mu^*  \tilde{f}(p_0,p_1) 
\,
\end{gather}
where we have defined
\begin{gather}
   \hat{p}_\mu = e^{i \frac{a p_\mu}{2}} \frac{2}{a} \sin \frac{a p_\mu}{2}
   \, .
\end{gather}

Introducing the collective bosonic and fermionic fields
\be
\begin{aligned}
   \Phi &= (  \text{Re}\,x ,  \text{Im}\,x , y^1 , \dots , y^5, \phi )^t \, , \\
   \Psi &= ( \theta_1 , \dots , \theta_4 , \theta^1 , \dots , \theta^4 , \eta_1 , \dots , \eta_4 , \eta^1 , \dots , \eta^4 )
   \, ,
\end{aligned}
\label{eq:collective-fields}
\ee
the free action~\eqref{S0} can be written in momentum space as
\begin{gather}
   S_0 = g \int_{-\pi/a}^{\pi/a} \frac{d^2p}{(2\pi)^2}
   \left\{ \tilde{\Phi}^t(-p) K_\text{B}(p) \tilde{\Phi}(p) + \tilde{\Psi}^t(-p) K_\text{F}(p) \tilde{\Psi}(p) \right\}
   \, ,
\end{gather}
where $K_\text{B}(p)$ is an $8 \times 8$ diagonal matrix for which the non-vanishing components given by
\begin{equation}
   K_B^{(n,n)}(p)
   =
   \begin{cases}
      c_+ |\hat{p}_0|^2 + c_- |\hat{p}_1|^2 + \frac{m^2}{2}
      \qquad & \text{if } n=1,2 \\
      c_+ |\hat{p}_0|^2 + |\hat{p}_1|^2
      \qquad & \text{if } n=3,\dots,7 \\
      c_+ |\hat{p}_0|^2 + |\hat{p}_1|^2 + m^2
      \qquad & \text{if } n=8
   \end{cases}
   \ ,
   \label{Eq KB}\\
\end{equation}
where we have defined the combinations
\begin{gather}
   c_\pm = b_\pm^2 \mp \frac{amb_\pm}{2}
   \ ,
\end{gather}
and $K_\text{F}(p)$ is an $16 \times 16$ matrix given by
\begin{align} 
   \label{Eq KF}
	K_F(p) =  \begin{pmatrix}
		0 &
      - \hat{p}_0^* I_{4 \times 4} &
      - \rho^6 \left( b_+ \hat{p}_1-\frac{im}{2} \right) &
      0
      \\
      - \hat{p}_0 I_{4 \times 4} &
      0 & 
      0 &
      \rho^6 \left( b_+ \hat{p}_1-\frac{im}{2} \right)
      \\
      \rho^6 \left( b_+ \hat{p}_1^*+\frac{im}{2} \right) &
      0 &
      0 &
      - \hat{p}_0^* I_{4 \times 4}
      \\
      0 &
      - \rho^6 \left( b_+ \hat{p}_1^*+\frac{im}{2} \right) &
      - \hat{p}_0 I_{4 \times 4} &
      0
	\end{pmatrix}\, ,
\end{align}
where we have used the identities $\rho^6 = (\rho^6)^* = - (\rho^6)^t = - {\rho^6}^\dag$ which are valid in the chosen representation (see  appendix~\ref{app:discretization}). The two matrices satisfy $K_B^t(p) = K_B(-p)$ and $K_F^t(p) = -K_F(-p)$.

Propagators in momentum space are defined by the entries of the inverse of these matrices up to trivial prefactors. The matrix $K_B(p)$ is diagonal and therefore easily inverted, while the matrix $K_F(p)$ is inverted by observing that
\begin{gather}
   K_F(p)^2 = \left( |\hat{p}_0|^2 + c_+ |\hat{p}_1|^2 + \frac{m^2}{4} \right) I_{16 \times 16} \ .
\end{gather}

The propagators are then easily calculated:
\begin{gather}
   \sum_{\sigma} a^2 \, e^{-ip \sigma}
   \langle x(\sigma)x^*(0) \rangle_0
   =
   \frac{1}{g}
   \frac{1}{
   c_+ |\hat{p}_0|^2 + c_- |\hat{p}_1|^2 + \frac{m^2}{2}
   }
   \ , \\
   \sum_{\sigma} a^2 \, e^{-ip \sigma}
   \langle y^a(\sigma)y^b(0) \rangle_0
   =
   \frac{1}{2g}
   \frac{\delta^{ab}}{
   c_+ |\hat{p}_0|^2 + c_- |\hat{p}_1|^2
   }
   \ , \\
   \sum_{\sigma} a^2 \, e^{-ip \sigma}
   \langle \phi(\sigma)\phi(0) \rangle_0
   =
   \frac{1}{2g}
   \frac{1}{
   c_+ |\hat{p}_0|^2 + |\hat{p}_1|^2 + m^2
   }
   \ , \\
   \sum_{\sigma} a^2 \, e^{-ip \sigma}
   \langle \theta_i(\sigma) \theta^j(0) \rangle_0
   =
   - \frac{1}{2g}
   \frac{
   \hat{p}_0^* \delta_i^j
   }{
   |\hat{p}_0|^2 + c_+ |\hat{p}_1|^2 + \frac{m^2}{4}
   }
   \ , \\
   \sum_{\sigma} a^2 \, e^{-ip \sigma}
   \langle\eta_i(\sigma)\eta^j(0)\rangle_0
   =
   - \frac{1}{2g}
   \frac{
   \hat{p}_0^* \delta_i^j
   }{
   |\hat{p}_0|^2 + c_+ |\hat{p}_1|^2 + \frac{m^2}{4}
   }
   \ , \\
   \sum_{\sigma} a^2 \, e^{-ip \sigma}
   \langle \theta_i (\sigma) \eta_j(0)\rangle_0
   =
   - \frac{1}{2g}
   \frac{
   \rho^6_{ij} \left( b_+ \hat{p}_1-\frac{im}{2} \right)
   }{
   |\hat{p}_0|^2 + c_+ |\hat{p}_1|^2 + \frac{m^2}{4}
   }
   \ , \\
   \sum_{\sigma} a^2 \, e^{-ip \sigma}
   \langle \theta^i(\sigma) \eta^j(0)\rangle_0
   =
   - \frac{1}{2g}
   \frac{
   ({\rho^6}^\dag)^{ij} \left( b_+ \hat{p}_1-\frac{im}{2} \right)
   }{
   |\hat{p}_0|^2 + c_+ |\hat{p}_1|^2 + \frac{m^2}{4}
   }
   \ ,
\end{gather}
where $\sigma$ is a shorthand notation for $(s,t)$. All other 2-point functions vanish. The denominators in the propagators reduce to a particular simple form if we choose $c_\pm = 1$, which is obtained for $b_\pm = \bar{b}_\pm$ with
\begin{gather}
   \bar{b}_\pm =  \sqrt{ 1 + \left( \frac{am}{4} \right)^2 } \pm \frac{am}{4}
   \ .
\end{gather}
As we will see in the following sections, this choice is also the correct one to
reproduce continuum results for the observables we consider in this paper.

Let us turn now to the interaction vertices. The expansion of $S_{\text{eff}}$ in powers of the fields $x$, $\phi$, $y$, $\theta$ and $\eta$ is fairly trivial except for terms involving the forward derivative of $z^M$. We observe that
\begin{eqnarray}
   \hat\partial_k z^M(x)
   = &&
   \frac{e^{\phi(x+a e_k)} u^M(x+a e_k) - e^{\phi(x)} u^M(x)}{a}
   \nonumber \\ = &&
   \frac{e^{\phi(x) + a \hat\partial_k\phi(x)} [ u^M(x) + a \hat\partial_k u^M(x) ] - e^{\phi(x)} u^M(x)}{a}
   \nonumber \\ = &&
   e^{\phi(x)} \left\{
   \hat\partial_k\phi(x) u^M(x)
   + \hat\partial_k u^M(x)
   + \frac{e^{a \hat\partial_k\phi(x)} - 1 - a \hat\partial_k\phi(x)}{a} u^M(x)
   \right\}
   \ .
\end{eqnarray}
The first two terms in the last expression survive in the naive $a \to 0$ limit,
while the third term takes into account the violation of the Leibniz and chain
rules at finite lattice spacing. By expanding the exponentials, one obtain terms
that have an arbitrary number of powers of $\hat\partial_k\phi(x)$ multiplied by
explicit powers of $a$. The number of derivatives and the number of factors of $a$
are related by dimensional analysis. Analogously one finds the following
formulae
\begin{gather}
   \hat\partial_k u^6(x)
   =
   \frac{
   - 2 y^c(x) \hat\partial_k y^c(x) - a [\hat\partial_k y^c(x)]^2
   }{2 \left\{ 1+\frac{1}{4}[ y^c(x) + a \hat\partial_k y^c(x) ]^2 \right\} \left\{ 1+\frac{1}{4}y(x)^2 \right\} }
   \ , \\
   \hat\partial_k u^b(x)
   =
   \frac{   
   - 2 y^c(x) \hat\partial_k y^c(x) - a [\hat\partial_k y^c(x)]^2
   }{
   4 \left\{ 1+\frac{1}{4}[y^c(x)+a\hat\partial_k y^c(x)]^2 \right\}
   \left\{ 1+\frac{1}{4}y(x)^2 \right\}
   } y^b(x)
   \ .
\end{gather}
Again, by expanding these expressions in $y$, one obtains terms an arbitrary
number of powers of $\hat\partial_k y^c(x)$ multiplied by explicit powers of
$a$. The number of derivatives and the number of factors of $a$ are related by
dimensional analysis.

By inspecting all terms one sees that, at each order in the perturbative
expansion, the interaction Lagrangian density in $x$ is a polynomial of the
fields $\Phi(x)$, $\Psi(x)$, their first derivatives $\hat\partial \Phi(x)$,
$\hat\partial \Psi(x)$, $\bar\partial \Psi(x)$, the lattice spacing $a$, and the
mass $m$. We will not write all vertices explicitly, however the following
observations will be useful later on.
\begin{itemize}
   \item Possible vertices are constrained by dimensional analysis: the boson
   fields have mass dimension 0, the fermion fields have mass dimension 1/2, the
   discrete derivatives and $m$ have mass dimension 1, and the lattice spacing
   has mass dimension -1, while vertices must have dimension 2.
   \item The considered action generates only terms that are proportional to
   $m^0$, $m^1$ or $m^2$.
   \item Vertices exist only with 0, 2, or 4 fermion fields.
   \item The considered action generates only terms that are proportional to
   $a^p$ with $p \ge -2$. In particular terms proportional to $a^{-2}$ are generated by the Jacobian determinant in eq.~\eqref{eq:Seff}.
\end{itemize}

	\section{Superficial degree of divergence}
\label{sec:powercounting}

The goal of this section is to show that the lattice-discretized theory is
non-renormalizable by power counting. To this end, we need to calculate the
superficial degree of divergence of the generic Feynman diagram.

Feynman integrands on the lattice are periodic functions in each component of
the momenta, with period $2\pi/a$. In particular they are not rational functions
as in the continuum, but rational trigonometric functions of the momenta. As
a consequence, the problem of establishing an appropriate power counting on the
lattice is subtler than in the continuum, and it was solved completely by by
Reisz \cite{Reisz1988} (see also e.g. \cite{Luscher1988,Capitani2003}).
Following Reisz, given a function $F$ of the loop momenta $q_{i=1,\dots,L}$, of
the external momenta $p_{i=1,\dots,E}$, and of the lattice spacing $a$, the
superficial degree of divergence $\deg F$ of the function $F$ is
defined by means from its asymptotic behaviour
\begin{equation}
F(\lambda q, p; m, a/ \lambda) \stackrel{\lambda \rightarrow \infty}{=} C_F \lambda^{\deg F}+O\left(\lambda^{\deg F-1}\right)
\,,
\label{asymptdeg}
\end{equation}
where $C=+_F \neq 0$. It is straightforward to show that $\deg (FG) = \deg F +
\deg G$ and $\deg (F^{-1}) = - \deg F$. As in the continuum, each loop integral
contributes with a superficial degree of divergence 2.

Denote by $\tilde{\Theta}_\alpha(p)$ the generic (bosonic or fermionic) field in momentum space. We consider here the connected $n$-point function in momentum space
\begin{gather}
   \langle \tilde{\Theta}_{\alpha_1}(p_1) \cdots \tilde{\Theta}_{\alpha_{E}}(p_{E}) \rangle_c
   =
   G_{\alpha}(p) \,
   (2\pi)^2 \sum_{\vec{n} \in \mathbb{Z}^2} \delta^2\left( \tfrac{2\pi}{a} \vec{n} - {\textstyle\sum_{i=1}^E} p_i \right)
   \ .
   \label{eq:npt-c}
\end{gather}
In this formula, we have used the fact that momentum conservation on the lattice
takes the form of a delta comb which accounts for the $2\pi/a$ periodicity
in momentum space. As in the continuum, the perturbative expansion of
$G_{\alpha}(p)$ has a representation in terms of a sum of Feynman integrals. We
introduce the amputated $n$-point function
\begin{gather}
   G^{\text{amp}}_{\alpha_1, \dots, \alpha_E}(p_1, \dots, p_E)
   =
   \sum_{\beta_1, \dots, \beta_E} G_{\beta_1, \dots, \beta_E}(p_1, \dots, p_E)
   \prod_{e=1}^E \left[ D^{-1}(p_e) \right]_{\alpha_e \beta_e}
   \ ,
   \label{eq:npt-amp}
\end{gather}
where $D(p)$ is the propagator matrix. $G^{\text{amp}}_{\alpha}(p)$ has a
representation in terms of a sum of Feynman integrals in which the external
lines have been amputated, and we will refer to them as external legs.

Since lines that do not belong to any loop do not contribute to the superficial
degree of divergence, we can restrict our analysis to diagrams that do not have
such lines, i.e. one-particle irreducible diagrams. Therefore consider the
generic one-particle irreducible Feynman diagram contributing to
$G^{\text{amp}}_{\alpha}(p)$, and let $A$ be the corresponding Feynman integral.
We will denote by $E_B$ and $E_F$ the number of external bosonic and fermionic
legs respectively, and by $I_B$ and $I_F$ the number of internal bosonic and
fermionic lines respectively. Let $l_{i=1,\dots,I}$ be the momentum flowing in
the $i$-th internal line (with $I=I_B+I_F$), and let $p_{e=1,\dots,E}$ be the
momentum flowing in the $e$-th external leg (with $E=E_B+E_F$). The Feynman
integral has the general form
\begin{equation}\label{eq:loop_integral}
A
=
\int_{-\frac{\pi}{a}}^{\frac{\pi}{a}} \frac{d^{2} q_{1}}{(2 \pi)^{2}} \cdots \int_{-\frac{\pi}{a}}^{\frac{\pi}{a}} \frac{d^{2} q_{L}}{(2 \pi)^{2}}
W(\hat{p}, \hat{l}; m, a) \prod_{i=1}^I D_i(\hat{l}_i; m, a)
\, ,
\end{equation}
where $D_i$ is the propagator associated to the $i$-th internal line, $W$ is
the product of all vertices, and $L$ is the number of loops.

The internal momentum $l_i$ can always be written as $l_i = P_i + Q_i$ where
$P_i$ is a linear combination of external momenta, and $Q_i$ is a linear
combination of loop momenta. Also, because of one-particle irreducibility, every
internal line belongs to a loop, so $Q_i$ is not identically zero. The
propagators are functions of $\hat{l}_i$, whose degree of divergence is
determined by looking at the asymptotic behaviour
\begin{equation}
   \hat{l}_i = e^{i \frac{a(P_i+Q_i)}{2}} \frac{2}{a} \sin \frac{a(P_i+Q_i)}{2}
   \xrightarrow{ \substack{q \to \lambda q \\ a \to a/\lambda} }
   e^{i \frac{a(P_i+\lambda Q_i)}{2\lambda}} \frac{2\lambda}{a} \sin \frac{a(P_i+ \lambda Q_i)}{2\lambda}
   = 
   \lambda \hat{Q}_i  + O(\lambda^0)
   \label{eq:li_asympt}
   \ .
\end{equation}
It follows easily that the degree of divergence of bosonic and fermionic
propagators are the same as in the continuum, i.e.
\begin{gather}
   \deg D_i = 
   \begin{cases}
      -2 \qquad & \text{if $i$ is a bosonic line} \\
      -1 \qquad & \text{if $i$ is a fermionic line}
   \end{cases}
   \ .
\end{gather}
The contribution to the degree of divergence of the Feynman integral of all
propagators is simply
\begin{gather}
   \deg \prod_i D_i =  \sum_i \deg D_i = -2 I_B - I_F \ .
\end{gather}

Each vertex contributes to the function $W$ with:
\begin{itemize}
   \item some integer power of $a$ and $m$, coming from the explicit dependence
   on these two parameters of the interaction Lagrangian, as discussed in
   Section~\ref{sec:expansion};
   
   \item a product of some $\hat{p}_e$ where $p_e$ is the momentum flowing in
   the $e$-th amputated external leg, coming from the discrete derivatives
   acting on fields in vertices which are Wick-contracted to external fields;
   
   \item a product of some $\hat{l}_i$ where $l_i$ is the momentum flowing in
   the $i$-th internal line, coming from the discrete derivatives acting on
   fields in vertices which are Wick-contracted to fields in other vertices or
   possibly the same vertex.

\end{itemize}
Notice that the degree of divergence of degree of divergence of $\hat{p}_e$ is
determined by the asymptotic behaviour
\begin{equation}
   \hat{p}_e = e^{i \frac{a p_e}{2}} \frac{2}{a} \sin \frac{a p_e}{2}
   \xrightarrow{ \substack{q \to \lambda q \\ a \to a/\lambda} }
   e^{i \frac{a p_e}{2\lambda}} \frac{2\lambda}{a} \sin \frac{a p_e}{2\lambda}
   = 
   \lambda^0 p_e + O(\lambda^{-1})
   \label{eq:pe_asympt}
   \ .
\end{equation}
Let $P_a$ and $P_m$ be the total number of $a$ and $m$ factors respectively, and let
$D_E$ and $D_I$ be the total number of discrete derivative acting on internal
and external lines respectively. Using eqs.~\eqref{eq:li_asympt}
and~\eqref{eq:pe_asympt} one derives the asymptotic behaviour
\begin{equation}
   W(\hat{p}, \hat{l}; m, a)
   \xrightarrow{ \substack{q \to \lambda q \\ a \to a/\lambda} }
   W( \lambda^0 p, \lambda \hat{q}; m, a/\lambda) \, [ 1 + O(\frac{1}{\lambda}) ]
   =
   \lambda^{D_I-P_a} W(p,\hat{q};m,a) \, [ 1 + O(\frac{1}{\lambda}) ]
   \ ,
\end{equation}
which implies
\begin{gather}
   \deg W = D_I-P_a \ .
\end{gather}

The superficial degree of divergence of the considered Feynman integral is given
by
\begin{gather}
   \deg A = -2L + \deg W + \sum_i \deg D_i
   =
   2 L + D_I - P_a - 2 I_B - I_F
   \ .
   \label{eq:degA-1}
\end{gather}

It is also interesting to calculate the mass dimension of the Feynman integral. Notice that
\begin{gather}
   \dim D_i = 
   \begin{cases}
      -2 \qquad & \text{if $i$ is a bosonic line} \\
      -1 \qquad & \text{if $i$ is a fermionic line}
   \end{cases}
   \ , \\
   \dim W = P_m - P_a + D_I + D_E
   \ ,
\end{gather}
which yields
\begin{gather}
   \dim A = 2L + \dim W + \sum_i \dim D_i = 2L + P_m - P_a + D_I + D_E - 2 I_B - I_F
   \ .
   \label{eq:dimA-1}
\end{gather}
On the other hand, $A$ is a term in the perturbative expansion of
$G^{\text{amp}}_{\alpha}(p)$. The mass dimension of the amputated $n$-point
function is calculated by observing that the mass dimension of a bosonic field
in Fourier space is -2, the mass dimension of a fermionic field in Fourier space
is -3/2, and the mass dimension of the momentum-conservation delta is -2. Using
eqs.~\eqref{eq:npt-c} and~\eqref{eq:npt-amp}, one obtains
\begin{align}
   \dim A &= \dim G^{\text{amp}}= \dim G + 2E_B + E_F = -2E_B -\frac{3}{2}E_F + 2 + 2E_B + E_F
  \nonumber\\ & =
   2 - \frac{1}{2} E_F
   \ .
   \label{eq:dimA-2}
\end{align}
Combining with eqs.~\eqref{eq:degA-1} and~\eqref{eq:dimA-1} we get our final
formula for the degree of divergence of $A$:
\begin{gather}
   \deg A = 2 - \frac{1}{2} E_F - P_m - D_E
   \ .
   \label{eq:dimA-3}
\end{gather}
This formula shows that the degree of divergence of one-particle irreducible
diagrams cannot be larger than $2$. However, since the degree of divergence does
not depend on the number of external bosonic legs, at any loop order the number
of divergent diagrams is infinite. This implies that one needs infinitely many
counterterms at any loop order to cancel the UV divergences. Without extra
constraints on the counterterms one would conclude that the theory is
non-renormalizable.

Since the Feynman diagrams with $P_a=0$ are the same ones that appear in a
continuum regularization, the same conclusion holds in this case. However it is
known that, in dimensional regularization, non-trivial cancellations of UV
divergences happen, effectively showing that the UV counterterms are highly
constrained. Even though some general argument exists for the UV finiteness of
the Green-Schwarz AdS$_5\times S^5$ string before any gauge fixing, we are not
aware of a complete derivation of such constraints in the gauge-fixed theory,
parametrized around the null-cusp background.

The question of whether a similar cancellation of UV divergences happens in the
lattice discretization is a legitimate one. We will see with a couple of
examples that unfortunately this does not work as well as in dimensional
regularization: a certain amount of fine-tuning is needed in order to reproduce
the continuum results.

	\section{Some calculations}\label{Sec: calculations}

\subsection{Cusp anomaly}\label{Subsec: cusp anomaly}

The partition function of the lattice-discretized theory is given by
\begin{gather}
   Z_{\text{cusp}} = \int d \Phi \, d \Psi \, e^{-S_{\text{eff}}}
\end{gather}
in terms of the collective fields $\Phi$ and $\Psi$ are defined in
eq.~\eqref{eq:collective-fields} and of the effective action $S_{\text{eff}}$ is
defined in eq.~\eqref{eq:Seff}. Since the logarithm of the partition function is
extensive, a complete calculation is performed by considering a finite
worldsheet with area $V_2$. At this point the integral defining the partition
function is finite and can be analytically calculated order by order in the
perturbative expansion. Finally one can define the free energy density in the
infinite-volume limit, i.e.
\begin{gather}
   \rho(g,m,a) = - \lim_{V_2 \to \infty} \frac{1}{V_2} \log
   Z_{\text{cusp}}(g,m,a,V_2) \ .
\end{gather}
As in every statistical system, the free energy is defined up to an additive
constant and only free-energy differences have physical meaning. It is also
interesting to notice that rescaling the integration measure in each lattice
point $d\Phi(s,t)d\Psi(s,t) \to \beta d\Phi(s,t)d\Psi(s,t)$ is equivalent to
rescaling $Z_{\text{cusp}} \to \beta^{\frac{V_2}{a^2}} Z_{\text{cusp}}$, i.e to
redefining $\rho \to \rho - a^{-2} \log \beta$.
This shows that quadratic divergences in the free energy are immaterial and can
be removed by rescaling the integration measure. We propose to identify the
following derivative of the free-energy density with the cusp anomalous dimension
difference of free-energy densities with the cusp anomalous dimension
\begin{gather}
   f(g,m,a) = \frac{4}{m} \frac{\partial}{\partial m} \rho(g,m,a) \ .
   \label{eq:f_rho}
\end{gather}
It is straightforward to show that this derivative coincides with the standard
definition in dimensional regularization, and it is also free from the
normalization ambiguity.\footnote{%
Notice that in eq.~\eqref{cusp_anom} the parameter $m$ is set equal to 1. In the continuum, the $m$ dependence can be be reintroduced by simple dimensional analysis, yielding $Z_{\text{cusp}} = e^{-\frac{f(g)}{8} m^2 V_{2}}$ and consequently $\rho = \frac{f(g)}{8} m^2$, which is indeed consistent with the definition~\eqref{eq:f_rho}.}

At leading order the path integral defining the partition function reduces to a
Gaussian integral, which yields
\begin{gather}
   \rho(g,m,a)
   =
   g \frac{m^2}{2} - \frac{4}{a^2} \log (2\pi) + \frac{1}{2} \int_{-\pi/a}^{\pi/a} \frac{d^2q}{(2\pi)^2} \log\left[ \frac{\det K_B(q)}{\det K_F(q)} \right]
   + O(g^{-1})
   \ .
\end{gather}
The determinants are calculated from the explicit expressions of $K_B$ and $K_F$
given in Section~\eqref{sec:expansion}, yielding
\begin{gather}
   \frac{\det K_B(q)}{\det K_F(q)}
   = 
   \tfrac{
   \left( c_+ |\hat{q}_0|^2 + c_- |\hat{q}_1|^2 + \frac{m^2}{2} \right)^2
   \left( c_+ |\hat{q}_0|^2 + |\hat{q}_1|^2 \right)^5
   \left( c_+ |\hat{q}_0|^2 + |\hat{q}_1|^2 + m^2 \right)
   }{
   \left( |\hat{q}_0|^2 + c_+ |\hat{q}_1|^2 + \frac{m^2}{4} \right)^8
   }
   \ .
\end{gather}

The calculation of $\rho$ and its small-$a$ expansion can be reduced to the
following general integral
\begin{gather}
   \int_{-\pi/a}^{\pi/a} \frac{d^2q}{(2\pi)^2} \log a^2 \left\{ \sum_i (1 + a \delta_i) |\hat{p}_i|^2 + M^2 \right\}
   \nonumber \\
   = \frac{1}{a^2} I_{-2}^{(0,0)}
   + \frac{\delta_1 + \delta_2}{2a}
   - \frac{\delta_1^2 + \delta_2^2}{4}
   + \frac{(\delta_1 - \delta_2)^2}{4\pi}
   - \frac{M^2}{4\pi} \log(aM)^2
   + M^2 I_0^{(0,0)} + O(a \log a)
   \ ,
\end{gather}
where $I_{-2}^{(0,0)}\simeq 1.166$ and $I_0^{(0,0)} \simeq 0.355$ are
numerical constants. The derivation of the above asymptotic expansion and the
precise definition of the constants are given in appendix~\ref{app:cusp}. By
using the above asymptotic expansion, with the convention $c_\pm = 1 + am \delta
c_\pm$, after a lengthy but straightforward calculation, one gets
\begin{eqnarray}
   \rho(g,m,a)
   & = &
   g \frac{m^2}{2} 
   - \frac{4 \log (2\pi)}{a^2}
   + \frac{m \delta c_-}{2a}
   - \frac{3 m^2  \log 2}{8\pi}
   - \frac{m^2 \delta c_-^2}{4}
   \nonumber \\ &&
   + \frac{m^2 \delta c_- (\delta c_- - 2 \delta c_+)}{4\pi}
   + O(a \log a) + O(g^{-1})
   \ ,
\end{eqnarray}
and, correspondingly, for the cusp anomaly:
\begin{gather}
   f(g,m,a)
   =
   4 g 
   + \frac{\delta c_-}{2a m}
   - \frac{3 \log 2}{\pi}
   - 2 \delta c_-^2
   + \frac{ 2 \delta c_- (\delta c_- - 2 \delta c_+) }{ \pi }
   + O(a \log a) + O(g^{-1})
   \ .
\end{gather}

Notice that with the naive choice $b_\pm=1$, which corresponds to $\delta c_\pm
= \mp 1/2$, the cusp anomaly contains a linear divergence. On the other hand,
with the special choice $b_\pm=\bar{b}_\pm$ which corresponds to $c_\pm = 1$ and
$\delta c_\pm = 0$, the linear divergence is canceled, and we obtain the same
result as in dimensional regularization:
\begin{gather}
   f(g,m,0)
   =
   4 g 
   - \frac{3\log 2}{\pi}
   + O(g^{-1})
   \ .
\end{gather}

\subsection{1-point functions}\label{Subsec: 1pt function}

Let us turn to the one-point functions of the perturbative fields. Notice that
$\langle x \rangle = 0$ because of the $U(1)$ symmetry, and $\langle y^a \rangle
= 0$ because of the $SO(5) \subset SO(6) \simeq SU(4)$ which leaves the
perturbative vacuum invariant. $\phi$ is the only field with a non-vanishing
one-point function, which has been calculated in dimensional
regularization~\cite{Giombi,Giombi:2010bj,Giombi2010}.  
This one-point function, as well as any $n$-point function of bare
fields, is not expected to be UV finite. In fact it is known that $\langle \phi
\rangle$ is UV divergent in dimensional regularization, and we will see that it
turns out to be UV divergent also in the lattice regularization. The interest in
this one-point function lies in the fact that it appears as a subdiagram in any
other $n$-point function, and ultimately its UV divergence contributes to any
physical observable. We will give an example of this mechanism in the next
subsection.

There are two classes of vertices contributing to the one-point function of
$\phi$: single-field vertices coming from the measure
\begin{gather}
   S_\phi 
   =
   -6 \sum_{s,t} \phi
   \ ,
\end{gather}
and three-field vertices coming from the action
\begin{eqnarray}
   S_{\phi \bullet \bullet}
   &=&
   g \sum_{s,t} a^2 \bigg\{
   -4 \phi \left| b_- \hat{\partial}_s x - \tfrac{m}{2} x \right|^2
   + c_+ \hat{\partial}_t \phi \hat{\partial}_t (\phi^2)
   + \hat{\partial}_s \phi \hat{\partial}_s \phi^2
   - 4 \phi (\hat{\partial}_s \phi)^2
   \nonumber \\ && \hspace{16mm}
   + 2 c_+ \hat{\partial}_t y^a \hat{\partial}_t ( \phi y^a )
   - c_+ \hat{\partial}_t \phi \hat{\partial}_t (y^2)
   + 2 \hat{\partial}_s y^a \hat{\partial}_s ( \phi y^a )
   - \hat{\partial}_s \phi \hat{\partial}_s (y^2)
   - 4 \phi (\hat{\partial}_s y^a)^2
   \nonumber \\ && \hspace{16mm}
   - 4 i \phi
   \left[
   \eta^i (\rho^6)_{ij} \left( b_+ \bar{\partial}_s \theta^j - \tfrac{m}{2} \theta^j \right)
   + \eta_i ({\rho^6}^\dag)^{ij} \left( b_+ \bar{\partial}_s \theta_j - \tfrac{m}{2} \theta_j \right)
   \right]
   \bigg\}
   \ .
\end{eqnarray}

Notice that the insertion of $S_\phi$ produces a tree-level diagram, while the
insertion of $S_{\phi \bullet \bullet}$ produces a one-loop diagram. However,
because of the mismatch in the power of $g$ in $S_\phi$ and $S_{\phi \bullet
\bullet}$, all these diagrams contribute to the same order in $g$, yielding
\begin{eqnarray}
   \langle \phi \rangle
   &=&
   \frac{3}{g m^2 a^2}
   + \frac{2}{g m^2}
   \int_{-\pi/a}^{\pi/a} \frac{d^2q}{(2\pi)^2}
   \frac{
   c_- |\hat{q}_1|^2 + \tfrac{m^2}{4}
   }{
   c_+ |\hat{q}_0|^2 + c_- |\hat{q}_1|^2 + \frac{m^2}{2}
   }
   \nonumber \\ &&
   - \frac{1}{2gm^2}
   \int_{-\pi/a}^{\pi/a} \frac{d^2q}{(2\pi)^2}
   \frac{
   c_+ |\hat{q}_0|^2 - |\hat{q}_1|^2
   }{
   c_+ |\hat{q}_0|^2 + |\hat{q}_1|^2 + m^2
   }
   - \frac{5}{2gm^2}
   \int_{-\pi/a}^{\pi/a} \frac{d^2q}{(2\pi)^2}
   \frac{
   c_+ |\hat{q}_0|^2 - |\hat{q}_1|^2
   }{
   c_+ |\hat{q}_0|^2 + |\hat{q}_1|^2
   }
   \nonumber \\ &&
   - \frac{8}{gm^2}
   \int_{-\pi/a}^{\pi/a} \frac{d^2q}{(2\pi)^2}
   \frac{
   c_+ |\hat{q}_1|^2 + \frac{m^2}{4}
   }{
   |\hat{q}_0|^2 + c_+ |\hat{q}_1|^2 + \frac{m^2}{4}
   }
   + O(g^{-2})
   \label{eq:phi-1}
   \ .
\end{eqnarray}

With the special choice $b_\pm=\bar{b}_\pm$, i.e. $c_\pm = 1$, one can use the
symmetry of the integrals under $p_0 \leftrightarrow p_1$ exchange to simplify
\begin{eqnarray}
   \langle \phi \rangle
   &=&
   - \frac{1}{g}
   \int_{-\pi/a}^{\pi/a} \frac{d^2q}{(2\pi)^2}
   \frac{
   1
   }{
   |\hat{q}|^2 + \frac{m^2}{4}
   } + O(g^{-2})
   \nonumber \\ &=&
   \frac{1}{g} \left\{
   \frac{1}{4\pi} \log \frac{(am)^2}{4}
   + \frac{1}{4\pi}
   - I_0^{(0,0)} + O(a \log a)
   \right\} + O(g^{-2})
   \ ,
   \label{eq:phi-2}
\end{eqnarray}
which is logarithmically divergent, as one can explicitly see by using the
asymptotic expansion given in appendix~\ref{app:1pt}. The definition of the
numerical constant $I_0^{(0,0)} \simeq 0.355$ is given in
appendix~\ref{app:cusp}. Notice that the measure, fermion-loop and $x$-loop
contributions are separately quadratically divergent, and the cancellation of
these divergences is highly non-trivial.

In the general case $c_\pm = 1 + (am) \delta c_\pm$ where $\delta c_\pm =
O(a^0)$, one can again use the asymptotic expansions given in
appendix~\ref{app:1pt}, and after a lengthy calculation one gets
\begin{gather}
   \langle \phi \rangle
   =
   \frac{1}{g} \bigg\{
   \frac{- 8 \delta c_+ + \delta c_-}{\pi a}
   + \frac{1}{4\pi} \log \frac{(am)^2}{4}
   \nonumber \\ \hspace{3cm}
   + \frac{1}{4\pi}
   - I_0^{(0,0)} 
   + \frac{8 \delta c_+^2 - \delta c_-^2}{2\pi}
   + O(a \log a) \bigg\} + O(g^{-2})
   \label{eq:phi-3}
   \ .
\end{gather}
Notice that the naive choice $b_\pm = 1$ corresponds to the choice $\delta c_\pm
= \mp 1/2$ which yields indeed a linear divergence for $\langle \phi \rangle$:
\begin{gather}
   \langle \phi \rangle
   =
   \frac{1}{g} \left\{ \frac{
   9
   }{2\pi a}
   + O(\log a) \right\} + O(g^{-2})
   \label{eq:phi-4}
   \ .
\end{gather}

	\subsection{2-point function}

We turn now to the two-point function of the field $x$, which we calculate at
one loop. We will use the two-point function to extract the dispersion relation
of the $x$ particle propagating on the worldsheet. In dimensional regularization
and at one loop \cite{Giombi:2010bj}, both the two-point function and the
dispersion relation turn out to be UV finite without any need of
renormalization. We will see that this is true also at one loop in lattice
perturbation theory, provided that one has chosen $c_\pm = 1$. The naive choice
$b_\pm = 1$ generates UV divergences in the dispersion relation. Whether these
divergences can be eliminated with a renormalization procedure is a valid
question.

\begin{figure}
   \centering
   \includegraphics[scale=0.6]{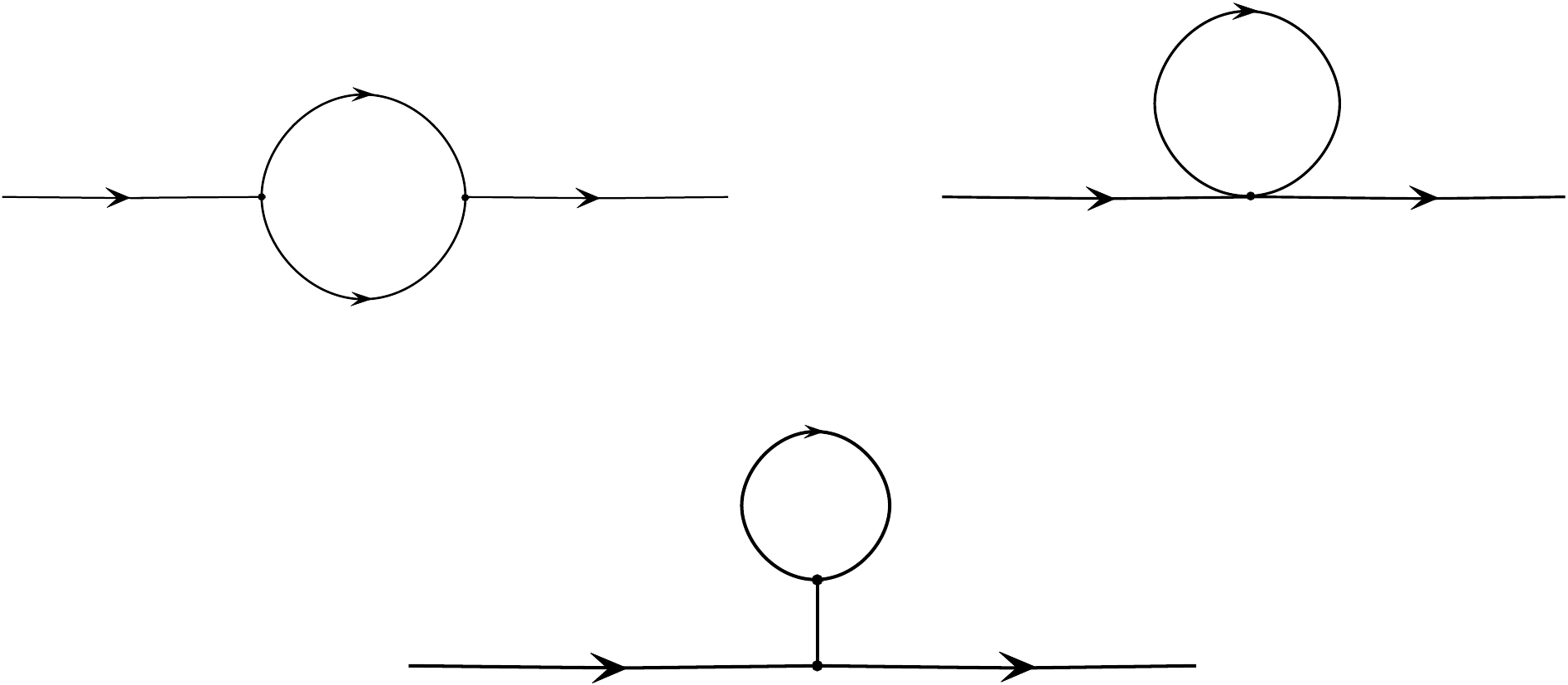}
   \caption{Topologies of diagrams contributing to the two point function at 1-loop.}
   \label{fig:conn_two_p}
\end{figure}

There are two classes of vertices contributing to the two-point function of
$x$ at one loop: three-field vertices
\begin{eqnarray}
   S_{x x^* \bullet}
   &=&
   g \sum_{s,t} a^2 \bigg\{ -4 \phi \left| b_- \hat{\partial}_s x - \tfrac{m}{2} x \right|^2
   \nonumber \\ && \hspace{18mm}
   + 2 \eta^i \rho^6_{ij} \eta^j \left( b_- \hat{\partial}_s x - \tfrac{m}{2} x \right)
   - 2 \eta_i ({\rho^6}^\dag)^{ij} \eta_j \left( b_- \hat{\partial}_s x^* - \tfrac{m}{2} x^* \right)
   \bigg\}
   \ ,
\end{eqnarray}
and four-field vertices
\begin{gather}
   S_{x x^* \bullet \bullet}
   =
   8 g \sum_{s,t} a^2 \phi^2 \left| b_- \hat{\partial}_s x - \tfrac{m}{2} x \right|^2
   \ ,
\end{gather}
combined to give Feynman diagrams with the three different topologies
illustrated in Fig.~\ref{fig:conn_two_p}. Notice that the tadpole contribution
will be proportional to $\langle \phi \rangle$.

On general grounds one sees that the two-point function has the following form
\begin{equation}\!\!\!\!\! 
   \langle \tilde{x}(p)x^*(0) \rangle
   =
   \frac{1}{g}
   \left\{
   c_+ |\hat{p}_0|^2 + c_- |\hat{p}_1|^2 + \frac{m^2}{2}
   + \frac{1}{g} \left( c_- |\hat{p}_1|^2 + \frac{m^2}{4} \right) \Pi_a(p) + O(g^{-2})
   \right\}^{-1}
   \ . \label{eq:2pt-x}
\end{equation}
The factor $\left(c_- |\hat{p}_1|^2 + \tfrac{m^2}{4} \right)$ comes from the
fact that, in all interaction vertices, $x$ always appears in the combination
$\left(b_- \hat{\partial}_s x - \tfrac{m}{2} x \right)$ or its complex
conjugate. The function $\Pi_a(p)$ has a representation in terms of amputated
Feynman diagrams and it is explicitly given by
\begin{eqnarray}
   \Pi_a(p)
   &=&
   - 4 g \langle \phi \rangle
   + 4 \int_{-\pi/a}^{\pi/a} \frac{d^2q}{(2\pi)^2}
   \frac{1}{c_+ |\hat{q}_0|^2 + |\hat{q}_1|^2 + m^2}
   \nonumber \\ &&
   - 8 \int_{-\pi/a}^{\pi/a} \frac{d^2q}{(2\pi)^2}
   \frac{c_- |\hat{q}_1|^2 + \frac{m^2}{4}}{c_+ |\hat{q}_0|^2 + c_- |\hat{q}_1|^2 + \frac{m^2}{2}}
   \frac{1}{
   c_+ |\widehat{p+q}_0|^2 + |\widehat{p+q}_1|^2 + m^2
   }
   \nonumber \\ &&
   - 8 \int_{-\pi/a}^{\pi/a} \frac{d^2q}{(2\pi)^2}
   \frac{\hat{q}_0}{|\hat{q}_0|^2 + c_+ |\hat{q}_1|^2 + \frac{m^2}{4}}
   \frac{\widehat{p+q}_0^*}{
   |\widehat{p+q}_0|^2 + c_+ |\widehat{p+q}_1|^2 + \frac{m^2}{4}
   }
   \ .
\end{eqnarray}
All integrals in the above formula are logarithmically divergent, while the term
proportional to $\langle \phi \rangle$ contains in general a linear divergence.
Up to terms that vanish in the $a \to 0$ limit, one can replace $c_\pm = 1$ in
the above integrals, obtaining the simpler expression
\begin{eqnarray}
   \Pi_a(p)
   &=&
   - 4 g \langle \phi \rangle
   + 4 \int_{-\pi/a}^{\pi/a} \frac{d^2q}{(2\pi)^2}
   \frac{1}{|\hat{q}|^2 + m^2}
   - 8 \int_{-\pi/a}^{\pi/a} \frac{d^2q}{(2\pi)^2}
   \frac{|\hat{q}_1|^2 + \frac{m^2}{4}}{|\hat{q}|^2 + \frac{m^2}{2}}
   \frac{1}{
   |\widehat{p+q}|^2 + m^2
   }
   \nonumber \\ &&
   - 8 \int_{-\pi/a}^{\pi/a} \frac{d^2q}{(2\pi)^2}
   \frac{\hat{q}_0}{|\hat{q}|^2 + \frac{m^2}{4}}
   \frac{\widehat{p+q}_0^*}{
   |\widehat{p+q}|^2 + \frac{m^2}{4}
   }
   + O(a \log a)
   \ .
\end{eqnarray}
As in the continuum, the leading divergence of the above integrals does not
depend on the external momentum, therefore the subtracted quantity $\Delta
\Pi_a(p) = \Pi_a(p) - \Pi_a(0)$ has a finite $a \to 0$ limit given by the corresponding continuum integrals, i.e.
\begin{eqnarray}
   \Delta \Pi_0(p)
   &=&
   - 8 \int_{-\infty}^\infty \frac{d^2q}{(2\pi)^2}
   \frac{q_1^2 + \frac{m^2}{4}}{q^2 + \frac{m^2}{2}}
   \left\{ \frac{1}{(p+q)^2 + m^2} - \frac{1}{q^2 + m^2} \right\}
   \nonumber \\ \label{Deltapi0} &&
   - 8 \int_{-\infty}^\infty \frac{d^2q}{(2\pi)^2}
   \frac{q_0}{|\hat{q}|^2 + \frac{m^2}{4}}
   \left\{
   \frac{p_0+q_0}{(p+q)^2 + \frac{m^2}{4}}
   - \frac{q_0}{q^2 + \frac{m^2}{4}}
   \right\}
   + O(a \log a)
   \ ,
\end{eqnarray}
while all the divergences are contained in
\begin{gather}
   \Pi_a(0)
   =
   - 4 g \langle \phi \rangle
   - 4 \int_{-\pi/a}^{\pi/a} \frac{d^2q}{(2\pi)^2}
   \frac{
   1
   }{
   |\hat{q}|^2 + \frac{m^2}{4}
   }
   + \frac{1}{\pi}
   + O(a \log a)
   \ ,
\end{gather}
where we have used the symmetry of the integrals under $p_0 \leftrightarrow p_1$
exchange to simplify them.

With the choice $c_\pm = 1$, using eq.~\eqref{eq:phi-2} one immediately sees that
all divergences cancel and $\Pi_0(0) = 1/\pi$. The two-point function is finite
in the continuum limit and 
\begin{gather}
   \lim_{a \to 0} \langle \tilde{x}(p)x^*(0) \rangle
   =
   \frac{1}{g}
   \left\{
   p^2 + \frac{m^2}{2}
   + \frac{1}{g} \left( p_1^2 + \frac{m^2}{4} \right) \Pi_0(p) + O(g^{-2})
   \right\}^{-1}
   \ ,
\end{gather}
The two-point function has a poles at $p_0 = \pm i E(p_1)$ for every value of
$p_1$, where $E(p_1)$ is the energy of a single excitation with the quantum
numbers of the field $x$, propagating on the worldsheet with momentum $p_1$. In
the continuum limit this is found to be
\begin{eqnarray}
   E(p_1)^2 &=& p_1^2 + \frac{m^2}{2}
   + \frac{1}{g} \left( p_1^2 + \frac{m^2}{4} \right) \Pi_0\left(\sqrt{p_1^2 + \frac{m^2}{2}},p_1\right) + O(g^{-2})
   \nonumber \\ &=&
   p_1^2 + \frac{m^2}{2}
   - \frac{1}{gm^2} \left( p_1^2 + \frac{m^2}{4} \right)^2
   + O(g^{-2})
   \ , \label{eq:disprelation}
\end{eqnarray}
where we have used the on-shell value of $\Pi_0$~\eqref{Pi0}. The obtained
dispersion relation coincides~\footnote{To compare with~\cite{Giombi:2010bj}, notice that one has to redefine the worldsheet coordinates, resulting in square masses of the fluctuations rescaled with a factor of $4$.} with the result in \cite{Giombi:2010bj}.

However in the general case $c_\pm = 1 + (am) \delta c_\pm$ where $\delta c_\pm
= O(a^0)$, $\Pi_a(0)$ and $E(p_1)$ inherit the linear divergence from $\langle
\phi \rangle$. Using eq.~\eqref{eq:phi-3} one obtains
\begin{gather}
   \Pi_a(0)
   =
   \frac{32 \delta c_+ - 4 \delta c_-}{\pi a}
   + \frac{1 - 16 \delta c_+^2 + 2 \delta c_-^2}{\pi}
   + O(a \log a)
   \ .
\end{gather}
For instance, for the naive choice $b_\pm = 1$, which corresponds to $\delta c_\pm
= \mp 1/2$, one obtains for the dispersion relation
\begin{gather}
   E(p_1)^2 = p_1^2 + \frac{m^2}{2}
   + \frac{1}{g} \left( p_1^2 + \frac{m^2}{4} \right)
   \left[ - \frac{18}{\pi a} + O(\log a) \right]
   + O(g^{-2})
   \ .
\end{gather}
It is interesting to notice that, once we have set $b_\pm = 1$, the divergence
in the dispersion relation cannot be eliminated by renormalizing the remaining
available parameters, i.e. $g$ and $m$. In other words, the choice $b_\pm = 1$
is not stable under renormalization. On the other hand, if one allows the
coefficients $b_\pm$ to be renormalized along with $m$ and $g$, then the
divergences in the dispersion relation are eliminated e.g. by choosing
\begin{gather}
   b_+ = 1 + \frac{1}{g_R} \frac{ \frac{am_R}{8} }{ 2 + \frac{am_R}{2} } \left( \Pi_a(0) - \frac{1}{\pi} \right) 
   \ , \\
   b_- = 1 - \frac{1}{g_R} \frac{ 1 + \frac{5am_R}{8} }{ 2 + \frac{am_R}{2} } \left( \Pi_a(0) - \frac{1}{\pi} \right) 
   \ , \\
   m^2 = m_R^2 \left[ 1 + \frac{1}{2g_R} \left( \Pi_a(0) - \frac{1}{\pi} \right) \right]
   \ , \\
   g = g_R \left[ 1 + O(g^{-1}) \right]
   \ .
\end{gather}
This choice yields a dispersion relation in the continuum limit of the same form
as eq.~\eqref{eq:disprelation}, except that the mass $m$ needs to be replaced by
its renormalized counterpart $m_R$. One could also see that the one-loop
renormalization of the coupling constant can be chosen in such a way that the
cusp anomaly be finite. With this discussion we do not want to imply that the
chosen lattice theory is renormalizable (we do not know this). However we
conclude that, if the lattice theory is renormalizable, then it is not
sufficient to renormalize $m$ and $g$, one also needs to introduce extra
coefficients in the action and either fine-tune their tree-level value, or
renormalize them.

\section*{Acknowledgements}
	
	We thank Edoardo Vescovi and Johannes Weber for discussions. The research of GB is funded from the European Union's Horizon 2020 research and innovation programme under the Marie Sklodowska-Curie ITN grant No 813942.    
	The research IC, and partially of GB, is funded by the Deutsche Forschungsgemeinschaft (DFG, German Research Foundation) - Projektnummer 417533893/GRK2575 "Rethinking Quantum Field Theory". 
	The research of VF is supported by the STFC grant ST/S005803/1, the European ITN grant No 813942 and from the Einstein Foundation Berlin through an Einstein Junior Fellowship.


	\appendix
	
	\section{$\rho$ matrices}
\label{app:discretization}

In the action \eqref{S_cusp_latt}  the matrices $\rho^M$ appear, which are off-diagonal blocks of the six-dimensional Dirac matrices
in chiral representation
\be 
\gamma^M\equiv \begin{pmatrix}
0  & {\rho^M}^\dagger   \\
 \rho^M   &  0 
\end{pmatrix}
=
\begin{pmatrix}
0  & (\rho^M)^{ij}   \\
(\rho^M)_{ij}   &  0 
\end{pmatrix}
\ee
\begin{align}
\rho_{ij}^M &=- \rho_{ji}^M\,, &
({\rho^M}^\dag)^{il}\rho_{lj}^N + ({\rho^N}^\dag)^{il}\rho_{lj}^M
&=2\delta^{MN}\delta_j^i \,.
\end{align}
The two off-diagonal blocks, carrying upper and lower indices respectively, are related by $(\rho^M)^{ij}=-(\rho^M_{ij})^*\equiv(\rho^M_{ji})^*$, so that the block with upper indices, $({\rho^M}^{\dagger})^{ij}$, is the conjugate transpose of the block with lower indices.
A possible explicit  representation is
\begin{equation}
\begin{aligned}
\rho^1_{ij}&=\left(\begin{matrix}0&1&0&0\\-1&0&0&0\\0&0&0&1\\0&0&-1&0
\end{matrix}\right)\,,&
\rho^2_{ij}&=\left(\begin{matrix}0&\mathrm{i}&0&0\\-\mathrm{i}&0&0&0\\0&0&0&-\mathrm{i}\\0&0&\mathrm{i}&0
\end{matrix}\right)\,,&
\rho^3_{ij}&=\left(\begin{matrix}0&0&0&1\\0&0&1&0\\0&-1&0&0\\-1&0&0&0
\end{matrix}\right)\,, \\
\rho^4_{ij}&=\left(\begin{matrix}0&0&0&-\mathrm{i}\\0&0&\mathrm{i}&0\\0&-\mathrm{i}&0&0\\ \mathrm{i}&0&0&0
\end{matrix}\right)\,,&
\rho^5_{ij}&=\left(\begin{matrix}0&0&\mathrm{i}&0\\0&0&0&\mathrm{i}\\-\mathrm{i}&0&0&0\\0&-\mathrm{i}&0&0
\end{matrix}\right)\,,&
\rho^6_{ij}&=\left(\begin{matrix}0&0&1&0\\0&0&0&-1\\-1&0&0&0\\0&1&0&0
\end{matrix}\right)\,.
\end{aligned}
\label{Eq:rho6}
\end{equation}
The $SO(6)$ generators are built out of the $\rho$-matrices via
\begin{equation}
\rho^{MN}{}^i{}_{ j} \equiv\frac{1}{2}[  ({\rho^M}^\dag)^{il}\rho_{lj}^N
- ({\rho^N}^\dag)^{il}\rho_{lj}^M ].
\end{equation}

\section{Asymptotic expansions of relevant integrals}
\label{app:asymptotic}

\subsection{Cusp anomaly}
\label{app:cusp}

We want to calculate the small-$a$ expansion of the following integral
\begin{eqnarray}
   F(a) &=& 
   \int_{-\pi/a}^{\pi/a} \frac{d^2 q}{(2\pi)^2}
   \log \left\{ a^2 \left[ \sum_i \alpha_i |\hat{p}_i|^2 + M^2 \right] \right\}
   \nonumber \\ &=&
   \frac{1}{a^2} \log (a M)^2
   + \int_{-\pi/a}^{\pi/a} \frac{d^2 q}{(2\pi)^2}
   \log \frac{ \sum_i \alpha_i |\hat{p}_i|^2 + M^2 }{ M^2 }
   \ . \label{eq:app:int1:F0}
\end{eqnarray}
Using the Schwinger-time representation of the logarithm, i.e.
\begin{gather}
   \log \frac{ \sum_i \alpha_i |\hat{p}_i|^2 + M^2 }{ M^2 }
   =
   - \int_0^\infty \frac{ds}{s} \, \left\{
   e^{- s \left[ a^2 \sum_i \alpha_i |\hat{q}_i|^2 + (aM)^2 \right] }
   - e^{- s (aM)^2 }
   \right\}
   \ ,
\end{gather}
and the change of variable $z = a q$, we obtain
\begin{gather}
   F(a)
   =
   \frac{1}{a^2} \log (a M)^2
   - \frac{1}{a^2} \int_0^\infty \frac{ds}{s} \, e^{-s (aM)^2} \left\{ K(\alpha_1 s) K(\alpha_2 s) - 1 \right\}
   \ , \label{eq:app:int1:F1}
\end{gather}
with the definition
\begin{gather}
   K(s)
   = 
   \int_{-\pi}^{\pi} \frac{d z}{2\pi}
   e^{- 4 s \sin^2 \frac{z}{2} }
   =
   \frac{1}{\sqrt{4\pi s}} + O(s^{-2})
   \ . \label{eq:app:int1:K}
\end{gather}
The function $K(s)$ is infinitely differentiable in $[0,\infty)$, and its
large-$s$ asymptotic behaviour is obtained by means of a standard saddle-point
analysis. We split the integral in eq.~\eqref{eq:app:int1:F1} in two regions,
and we write
\begin{eqnarray}
   F(a)
   &=&
   \frac{1}{a^2} \log (a M)^2
   - \frac{1}{a^2} \int_0^1 ds \, e^{-s (aM)^2} \frac{ K(\alpha_1 s) K(\alpha_2 s) - 1 }{s}
   \nonumber \\ &&
   - \frac{1}{a^2} \int_1^\infty \frac{ds}{s} \, e^{-s (aM)^2} K(\alpha_1 s) K(\alpha_2 s)
   + \frac{1}{a^2} \Gamma(0,(aM)^2)
   \ , \label{eq:app:int1:F2}
\end{eqnarray}
We also introduce the auxiliary function
\begin{gather}
   G(s)
   =
   \int_s^\infty \frac{d\sigma}{\sigma} \, K(\alpha_1 \sigma) K(\alpha_2 \sigma)
   =
   \frac{1}{4\pi \sqrt{\alpha_1 \alpha_2} s} + O(s^{-1})
   \ . \label{eq:app:int1:G}
\end{gather}
Thanks to the asymptotic behaviour~\eqref{eq:app:int1:K}, the above integral is
finite and its large-$s$ asymptotic behaviour easily follows. In terms of the
auxiliary function, and after integration by parts, the integral in the
large-$s$ region in eq.~\eqref{eq:app:int1:F2} reads
\begin{gather}
   - \frac{1}{a^2} \int_1^\infty \frac{ds}{s} \, e^{-s (aM)^2} K(\alpha_1 s) K(\alpha_2 s)
   =
   \frac{1}{a^2} \int_1^\infty ds \, e^{-s (aM)^2} G'(s)
   = \nonumber \\ =
   - \frac{1}{a^2} G(1)
   + M^2 \int_1^\infty ds \, e^{-s (aM)^2} G(s)
   = \nonumber \\ =
   - \frac{e^{-(aM)^2}}{a^2} G(1)
   + \frac{M^2}{4\pi \sqrt{\alpha_1 \alpha_2}} \Gamma(0,(aM)^2)
   \nonumber \\ \hspace{3cm}
   + M^2 \int_1^\infty ds \, e^{-s (aM)^2} \left\{ G(s) - \frac{1}{4\pi \sqrt{\alpha_1 \alpha_2} s} \right\}
   \ . \label{eq:app:int1:F3}
\end{gather}
In the last step we have added and subtracted the leading asymptotic
behaviour~\eqref{eq:app:int1:G}. Bringing together
eqs.~\eqref{eq:app:int1:F2} and \eqref{eq:app:int1:F3}, and expanding for
small $a$, we obtain
\begin{gather}
   F(a)
   =
   \frac{1}{a^2} I_{-2}(\alpha)
   - \frac{M^2}{4\pi \sqrt{\alpha_1 \alpha_2}} \log(aM)^2
   + M^2 I_0(\alpha) + O(a^2 \log a)
   \ , \label{eq:app:int1:F4}
\end{gather}
with the definitions
\begin{gather}
   I_{-2}(\alpha) = - \gamma - \int_0^1 ds \, \frac{ K(\alpha_1 s) K(\alpha_2 s) - 1 }{s}
   - G(1)
   \ , \\
   I_0(\alpha) =
   - \frac{\gamma}{4\pi \sqrt{\alpha_1 \alpha_2}}
   + \int_0^1 ds \, K(\alpha_1 s) K(\alpha_2 s)
   + G(1)
   + \int_1^\infty ds \, \left\{ G(s) - \frac{1}{4\pi \sqrt{\alpha_1 \alpha_2} s} \right\}
   \ .
\end{gather}
By using the definition of $G(s)$ and after some straightforward algebra, one
also obtains the representation
\begin{gather}
     I_{-2}(\alpha) = - \gamma - \int_0^1 ds \, \frac{ K(\alpha_1 s) K(\alpha_2 s) - 1 }{s}
   - \int_1^\infty \frac{ds}{s} \, K(\alpha_1 s) K(\alpha_2 s)
   \ , \\
      I_0(\alpha) =
   \frac{1-\gamma}{4\pi \sqrt{\alpha_1 \alpha_2}}
   + \int_0^1 ds \, K(\alpha_1 s) K(\alpha_2 s)
   + \int_1^\infty ds \left\{ K(\alpha_1 s) K(\alpha_2 s) - \frac{1}{4\pi \sqrt{\alpha_1 \alpha_2} s} \right\}\,.
\end{gather}
We are interested in eq.~\eqref{eq:app:int1:F4} with the special choice
$\alpha_i = 1 + a \delta_i$. By Taylor expanding eq.~\eqref{eq:app:int1:F4} in
$a \delta_i$, we obtain
\begin{eqnarray}
   F(a)
   &=&
   \frac{1}{a^2} I_{-2}^{(0,0)}
   + \frac{\delta_1 + \delta_2}{a} I_{-2}^{(1,0)}
   + \frac{\delta_1^2 + \delta_2^2}{2} I_{-2}^{(2,0)}
   + \delta_1 \delta_2 I_{-2}^{(1,1)}
   \nonumber \\ &&
   - \frac{M^2}{4\pi} \log(aM)^2
   + M^2 I_0(1,1) + O(a \log a)
   \ , \label{eq:app:int1:F5}
\end{eqnarray}
with the definitions
\begin{gather}
   I_{-2}^{(0,0)} = I_{-2}(1,1)
   =
   - \gamma - \int_0^1 ds \, \frac{ [K(s)]^2 - 1 }{s}
   - \int_1^\infty \frac{ds}{s} \, [K(s)]^2
   \ , \\
   I_{-2}^{(1,0)} = \frac{\partial I_{-2}}{\partial \alpha_1} (1,1)
   =
   - \int_0^\infty ds \, K'(s) K(s)
   =
   - \frac{1}{2} \int_0^\infty ds \, \frac{d}{ds} [K(s)]^2
   =
   \frac{1}{2}
   \ , \\
   I_{-2}^{(1,1)} = \frac{\partial^2 I_{-2}}{\partial \alpha_1 \partial \alpha_2} (1,1)
   =
   - \int_0^\infty ds \, s [ K'(s) ]^2
   =
   -\frac{1}{2\pi}
   \ , \\
   I_{-2}^{(2,0)} = \frac{\partial^2 I_{-2}}{\partial \alpha_1^2} (1,1)
   =
   - \int_0^\infty ds \, s K''(s) K(s)
   =
   \int_0^\infty ds \, K'(s) \frac{d}{ds} [ s K(s) ]
   \nonumber \\ \hspace{2cm} =
   \int_0^\infty ds \, [ K'(s) ]^2
   + \int_0^\infty ds \, K'(s) K(s)
   =
   \frac{1}{2\pi}
   - \frac{1}{2}
   \ , \\
   I_0^{(0,0)} = I_0(1,1) =
   \frac{1-\gamma}{4\pi}
   + \int_0^1 ds \, [K(s)]^2
   + \int_1^\infty ds \, \left\{ [K(s)]^2 - \frac{1}{4\pi s} \right\}
   \ .
\end{gather}
The unknowns integrals can be calculated numerically, yielding
$I_{-2}^{(0,0)}\simeq 1.166$ and $I_0^{(0,0)} \simeq 0.355$.

\subsection{1-point function}
\label{app:1pt}

By taking the derivative with respect to $M^2$ of both sides of
eq.~\eqref{eq:app:int1:F4}, and by using the definition~\eqref{eq:app:int1:F0},
we obtain
\begin{eqnarray}
   \int_{-\pi/a}^{\pi/a} \frac{d^2 q}{(2\pi)^2}
   \frac{1}{ \sum_i \alpha_i |\hat{p}_i|^2 + M^2 }
   &=&
   - \frac{1}{4\pi \sqrt{\alpha_1 \alpha_2}} \log(aM)^2
   \nonumber \\ &&
   - \frac{1}{4\pi \sqrt{\alpha_1 \alpha_2}}
   + I_0(\alpha) + O(a^2 \log a)
   \ , \label{eq:app:int2:as1}
\end{eqnarray}
Specializing to $\alpha_i = 1 + a \delta_i$ and Taylor-expanding in $a
\delta_i$, we obtain
\begin{eqnarray}
   \int_{-\pi/a}^{\pi/a} \frac{d^2 q}{(2\pi)^2}
   \frac{1}{ \sum_i (1 + a \delta_i) |\hat{p}_i|^2 + M^2 }
   &=&
   - \frac{1}{4\pi} \log(aM)^2
   - \frac{1}{4\pi}
   + I_0^{(0,0)} + O(a \log a)
   \ . \label{eq:app:int1:as2}
\end{eqnarray}
By applying the differential operator $\sum_i \beta_i \frac{\partial}{\partial
\alpha_i}$ to both sides of eq.~\eqref{eq:app:int1:F4}, and by using the
definition~\eqref{eq:app:int1:F0}, we obtain
\begin{eqnarray}
   \int_{-\pi/a}^{\pi/a} \frac{d^2 q}{(2\pi)^2}
   \frac{ \sum_i \beta_i |\hat{p}_i|^2 }{ \sum_i \alpha_i |\hat{p}_i|^2 + M^2 }
   &=&
   \frac{1}{a^2} \sum_i \beta_i \frac{\partial I_{-2}}{\partial
   \alpha_i}(\alpha)
   + \frac{M^2 (\beta_1 \alpha_2 + \beta_2 \alpha_1)}{8\pi (\alpha_1 \alpha_2)^{3/2}} \log(aM)^2
   \nonumber \\ &&
   + M^2 \sum_i \beta_i \frac{\partial I_0}{\partial
   \alpha_i}(\alpha) + O(a^2 \log a)
   \ . \label{eq:app:int1:as3}
\end{eqnarray}
Specializing to $\alpha_i = 1 + a \delta_i$ and Taylor-expanding in $a
\delta_i$, we obtain
\begin{gather}
   \int_{-\pi/a}^{\pi/a} \frac{d^2 q}{(2\pi)^2}
   \frac{ \sum_i \beta_i |\hat{p}_i|^2 }{ \sum_i (1 + a \delta_i) |\hat{p}_i|^2 + M^2 }
   =
   \frac{\beta_1+\beta_2}{a^2} I_{-2}^{(1,0)}
   + \frac{\beta_1\delta_1+\beta_2\delta_2}{a} I_{-2}^{(2,0)}
   \nonumber \\ \hspace{15mm}
   + \frac{\beta_1\delta_2+\beta_2\delta_1}{a} I_{-2}^{(1,1)}
   + \frac{\beta_1 \delta_1^2 + \beta_2 \delta_2^2}{2} I_{-2}^{(3,0)}
   + \frac{\beta_1 \delta_2^2  + \beta_2 \delta_1^2 + 2 ( \beta_1 + \beta_2) \delta_1 \delta_2}{2} I_{-2}^{(2,1)}
   \nonumber \\ \hspace{15mm}
   + \frac{M^2 (\beta_1 + \beta_2)}{8\pi} \log(aM)^2
   + M^2 (\beta_1+\beta_2) I_0^{(1,0)} + O(a \log a)
   \ ,
\end{gather}
with the following definitions
\begin{gather}
   I_{-2}^{(2,1)}(\alpha) = \frac{\partial^3 I_{-2}}{\partial \alpha_1^2 \partial \alpha_2}(1,1) 
   =
   - \int_0^\infty ds \, s^2 K''(s) K'(s)
   =
   - \frac{1}{2} \int_0^\infty ds \, s^2 \frac{d}{ds} [K'(s)]^2
   \nonumber \\ \hspace{2cm} =
   \int_0^\infty ds \, s [K'(s)]^2
   =
   \frac{1}{2\pi}
   \ , \\
   I_{-2}^{(3,0)}(\alpha) = \frac{\partial^3 I_{-2}}{\partial \alpha_1^2}(1,1)
   =
   - \int_0^\infty ds \, s^2 K'''(s) K(s)
   =
   \int_0^\infty ds \, K''(s) \frac{d}{ds} [ s^2 K(s) ]
   \nonumber \\ \hspace{2cm} =
   2 \int_0^\infty ds \, s K''(s) K(s)
   + \int_0^\infty ds \, s^2 K''(s) K'(s)
   \nonumber \\ \hspace{2cm} =
   - 2 \int_0^\infty ds \, K'(s) \frac{d}{ds} [ s K(s) ]
   - \frac{1}{2\pi}
   \nonumber \\ \hspace{2cm} =
   - 2 \int_0^\infty ds \, K'(s) K(s)
   - 2 \int_0^\infty ds \, s [K'(s)]^2
   - \frac{1}{2\pi}
   =
   1 - \frac{3}{2\pi}
   \ , \\
   I_0^{(1,0)} =
   - \frac{1-\gamma}{8\pi}
   + \int_0^1 ds \, s K'(s) K(s)
   + \int_1^\infty ds \left\{ s K'(s) K(s) + \frac{1}{8\pi s} \right\}
   \nonumber \\ \hspace{2cm} =
   - \frac{1-\gamma}{8\pi}
   + \frac{1}{2} \int_0^1 ds \, s \frac{d}{ds} [K(s)]^2
   + \frac{1}{2} \int_1^\infty ds \, s \frac{d}{ds} \left\{ [K(s)]^2 - \frac{1}{4\pi s} \right\}
   \nonumber \\ \hspace{2cm} =
   \frac{\gamma}{8\pi}
   - \frac{1}{2} \int_0^1 ds \, [K(s)]^2
   - \frac{1}{2} \int_1^\infty ds \, \left\{ [K(s)]^2 - \frac{1}{4\pi s} \right\}
   \nonumber \\ \hspace{2cm} =
   -\frac{1}{2} I_0^{(0,0)} + \frac{1}{8\pi}
   \ ,
\end{gather}
in addition to the definitions given in the previous subsection.

\subsection{Calculation of $\Delta \Pi_0$ }
\label{app:2pt}

The finite, continuum integral defined in the main text for the 2-point function in equation~\ref{Deltapi0} can be rewritten as the dimensionless integral
\begin{align}
	\Delta \Pi_0(p) =& -8\int \frac{d^2q}{(2\pi)^2}\left(\frac{q_1^2+1}{(q^2+2)((\tilde{p}+q)^2+4)}-\frac{1}{2}\frac{1}{q^2+4}\right)\nonumber \\
	& -8\int \frac{d^2q}{(2\pi)^2}\left(\frac{q_0^2+\tilde{p}_0 q_0}{(q^2+1)((\tilde{p}+q)^2+1)}-\frac{1}{2}\frac{1}{q^2+1}\right)-\frac{1}{\pi}
\end{align}
by rescaling the momenta $\tilde{p}=\frac{m}{2}p$ and manipulating the integrals. Using standard Feynman parametrisation, this can be recast as the integral
\begin{align}
	\Delta \Pi_0(p) &= \frac{-1}{\pi}\int_0^1 dx\left( \frac{(p_0^2-p_1^2)x^2+2\tilde{p}_1^2 x-(\tilde{p}^2+1)}{1+\tilde{p}^2x(1-x)}+\frac{(\tilde{p}_1^2-\tilde{p}_0^2)(1-x)^2}{4-2x+\tilde{p}^2x(1-x)}\right)-\frac{1}{\pi}.
\end{align}
Reverting to $p=\frac{2}{m}\tilde{p}$ and evaluating this at the on-shell value, we obtain
\begin{align}
	\Delta \Pi_0\left(p;p^2=\frac{m^2}{2}\right)&= \frac{-1}{m^2}\left(p_1^2+\frac{m^2}{4}\right)-\frac{1}{\pi}
\end{align}
Notice that for the choice $c_\pm=1$ where $ \Pi_0(0)=\frac{1}{\pi}$, we recover the continuum limit found in \cite{Giombi:2010bj},
\begin{align}\label{Pi0}
	\Pi_0\left(p;p^2=\frac{m^2}{2}\right)|_{c_\pm=1}&= \frac{-1}{m^2}\left(p_1^2+\frac{m^2}{4}\right)
\end{align}

	\bibliographystyle{nb}
	\bibliography{Ref_strings_lattice.bib}
	
\end{document}